\documentclass[twocolumn,showpacs,amssymb,nobibnotes,aps,floats,prb]{revtex4-1}
\usepackage{textcomp,amssymb,graphicx}
\usepackage{hyperref}
\usepackage{amsbsy}
\usepackage{amsmath}
\usepackage[dvips]{color}
\usepackage{graphicx}
\usepackage{float}
\date{}
\def \s{\sigma}
\def \sb{\bar\sigma}
\def \be{\begin{eqnarray}}
\def \ee{\end{eqnarray}}

\def \d{\dagger}
\def \sd{\sum\limits}
\def \sp{\!\!\!}
\def \spb{\sp\sp}
\begin{document}
\title{Phase diagram of Hubbard-Holstein model on 4-leg tube system at quarter-filling}
\author{Sahinur Reja${^1}$, Satoshi Nishimoto$^{2,3}$}
\affiliation{${^1}$Physics Department, Indiana University Bloomington, USA 47405 }
\affiliation{${^2}$Department of Physics, TU Dresden, 01069 Dresden, Germany}
\affiliation{${^3}$IFW Dresden, Helmholtzstrae 20, 01069 Dresden, Germany }
\date{\today}
\begin{abstract}
{We derive an effective electronic Hamiltonian for square lattice Hubbard-Holstein model (HHM)
in the strong electron-electron (e-e) and electron-phonon (e-ph) 
coupling regime and under non-adiabatic conditions ($t/\omega_0 \leq 1$), $t$ and 
$\omega_0$ being the electron hopping and phonon frequency respectively. Using Density Matrix Renormalization Group method, we simulate this effective electronic model 
on $4-$Leg cylinder system at quarter-filling and present a phase diagram in $g-U$ plane where $g$ and $U$ are being the e-ph coupling constant and Hubbard on-site interaction respectively. For larger $g$, we find cluster of spins i.e. phase separation (PS) gives way to a charge density wave (CDW) phase made of NN singlets which abruptly goes to another CDW phase as we increase $U$. But for smaller $g$, we find a metallic phase sandwiched between PS and singlet CDW phase. This phase is characterized by vanishing charge gap but finite spin gap -- suggesting a singlet superconducting phase.
}
\end{abstract}
\maketitle
\section{Introduction}
More than one type of interactions typically manifests a variety of phases such as diagonal long range orders [such as charge density
wave (CDW) and spin density wave (SDW)] 
and off-diagonal long range orders (such as superfluid and superconducting states) of which some cooperate and some compete. The study of coexistence and competition between
these electronic phases is a subject
of immense ongoing focus. In particular, the coexistence of
CDW and superconductivity/superfluidity in layered dichalcogenides (e.g., 2H-$\rm TaSe_2$,
2H-$\rm TaS_2$, and $\rm NbSe_2$) \cite{withers}, helium-4 \cite{chan},
bismuthates (e.g., $\rm BaBiO_3$ doped with $\rm K$ or $\rm P$) \cite{blanton}, quarter-filled organic materials \cite{mori,mckenzie},
non-iron based pnictides (e.g., $\rm SrPt_2As_2$)
{\cite{kudo}, quasi-one-dimensional (1D)
trichalcogenide $\rm NbSe_3$ \cite{chaikin} and doped spin ladder cuprate
$\rm Sr_{14}Cu_{24}O_{41}$ \cite{abbamonte}, and recently in optical lattice system with effective long-range interactions{\cite{opti_latt}} etc. 

Elecron-phonon (e-ph) coupling along with usual electron-electron (e-e) interaction plays an important role in condensed matter systems such as cuprates \cite{photoem1,photoem3}
and manganites \cite{lanzara2,pbl,boothroyd} and molecular solids such as fullerides \cite{fullarene3}. The interplay of e-e and e-ph interactions in these correlated systems gives rise to the competition/coexistence of various phase such as superconductivity, CDW, SDW etc. 

The simplest model to study the co-occurring effects of e-e and e-ph
interactions is the following well known Hubbard-Holstein model (HHM) \cite{sryspbl1}
\be
H_{hh} \!&=&\! -t\sum_{j,\delta,\sigma}\left(c^{\dagger}_{j\sigma}c_{j+\delta,\sigma}+ {\rm H.c.}
 \right)
+\omega_0\sum_{j}a_{j}^{\dagger}a_{j}\nonumber\\
\spb && +g\omega_0\sum_{j\sigma}n_{j\sigma}(a_{j}+a_{j}^{\dagger})
+U\sum_{j}n_{j\uparrow}n_{j\downarrow}  ,
\label{ai1}
\ee
where  $c_{j\sigma}^{\dagger}$  is
the fermionic creation operator for itinerant spin-$\sigma$ electrons at site $j$ with hopping integral $t$
and  number operator $n_{j\sigma} \equiv c_{j\sigma}^{\dagger}c_{j\sigma}$. Here $\delta=(\hat{x},\hat{y})$ with unit lattice parameter represents the nearest neighbors for square lattice which we consider for our calculations;
 $a_{j}^{\dagger}$ is the corresponding bosonic creation operator characterized by a 
dispersionless phonon
frequency $\omega_0$, with $U$ and $g$ representing the 
strengths of onsite e-e and e-ph interactions
respectively.

{The Hubbard-Holstein model
has been extensively studied (in one-, two-, and infinite-dimensions and at various fillings)
by employing various approaches such as
exact diagonalization \cite{exdiag1,exdiag2,exdiag4},
density matrix renormalization group (DMRG)\cite{dmrg,dmrg2},
quantum Monte Carlo (QMC) \cite{qmc1,qmc3,qmc4,qmc5,qmc6,qmc7},
dynamical mean field theory (DMFT)
\cite{dmft1,dmft2,dmft3,dmft4,dmft5,dmft6,dmft7,dmft8,dmft9},
semi-analytical slave
boson approximations \cite{slave_b1,slave_b2,slave_b3,slave_b4,slave_b5},
large-N expansion \cite{largeN2}, variational methods
based on Lang-Firsov transformation \cite{lf1,lf2},  Gutzwiller
approximation\cite{GA1,GA2}, and cluster approximation \cite{vca}.}

However, the  study of the subtle interplay of
e-e and e-ph interaction effects
in low-dimensional systems, such as
conjugated polymers, charge transfer salts, inorganic
spin-Peierls compounds, halogen-bridged transition
metal complexes, ferroelectric perovskites, or organic
superconductors \cite{quasi1d1,quasi1d2,quasi1d3,quasi1d4}, has attracted much attention. Apart from the superconductors, e-ph coupling in quasi-1D materials sometimes can drive the electrons to be insulating with a CDW by Peierls transition.

{
In our earlier work\cite{sryspbl1, sryspbl2}, in strong e-e and e-ph coupling regime,
we derived an effective electronic Hamiltonian using a controlled analytic approach
that takes into account dynamical quantum phonons. It was shown that
the e-ph interaction generates nearest-neighbor (NN) repulsion which competes with NN spin antiferromagnetic (AF) interactions produced by e-e interactions. This competition stabilizes a correlated NN singlet phase for intermediate e-e and e-ph coupling which was shown be a superfluid at all fillings (less than one-half) other that one-third where it is a CDW.  
}

In this paper, we study the HHM on a 4-Leg tube system at quarter-filling using DMRG method which is very effective in studying ground-state properties of quasi-1D systems with short-range interactions\cite{dmrg_white}. 
We show that NN singlet phase we uncovered for 1D HHM model\cite{sryspbl1, sryspbl2} still survives, but these singlets arrange themselves to form a CDW at quarter filling with finite charge and spin gap. This phase is shown to be stabilized between phase separation at smaller $U$ and a CDW phase at larger $U$. At smaller e-e and e-ph coupling, we find a metallic phase in the vicinity of NN singlet-CDW phase and phase separation with vanishing charge gap, but with finite spin gap -- suggesting a singlet superconducting phase. 

{The paper is organized as follows: in Sec. II we briefly derive the effective 
electronic Hamiltonian and explain the
various interaction terms and hopping terms. We also briefly mention the details of DMRG simulations.
In Sec. III, we present a phase diagram in $g-U/t$ plane mentioning different stable phases. 
Next, in Sec. IV we describe how we determine the different phase boundaries using DMRG simulations. 
For this we calculate charge gap, spin gap and different order parameters to identify various phases. 
Finally we conclude in Sec V.}


\begin{figure}[htp]
\centering
\includegraphics[width=.99\linewidth]{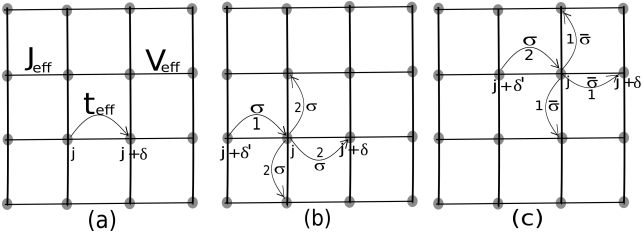}
\caption{
(a) The effective NN terms in the hamiltonian, (b) Longer range $\sigma$-spin hopping from site $j+\delta^{'}$ to site $j$ and then to $j+\delta$ with $\delta, \delta^{'}=\pm\hat{x},\pm\hat{y}$ as appropriate to avoid double counting. Here as shown for the case of $\delta^{'}=-\hat{x}$, we can have $\delta=\hat{x},\pm\hat{y}$. 
(c) Similarly $\delta^{'}=-\hat{x}$, and $\delta=\hat{x},\pm\hat{y}$ for the $\sigma\bar{\sigma}$ pair hopping where $\bar{\sigma}$-spin first hops from site $j$ to $j+\delta$ and then opposite spin $\sigma$ hops from site $j+\delta^{'}$ to $j$. The number $1(2)$ represents the first(second) hopping process.
}
\label{fig_model}
\end{figure}


\section{Effective HHM Hamiltonian}
Here we briefly outline the procedure to get the effective electronic 
Hubbard-Holstein Hamiltonian
(with more details being provided in Ref. \onlinecite{sryspbl1,pankaj_ys,amrita_ys}).
This approach involves a
 Lang-Firsov (LF) transformation \cite{LF_transf}
$H^{LF}_{hh}=e^{T}H_{hh}e^{-T}$
where $T=-g\sum_{j\sigma}n_{j\sigma}(a_{j}-a_{j}^{\dagger})$
and get the following LF transformed
Hamiltonian: \\
\be
H^{LF}_{hh}&=& -t\sum_{j\delta\sigma}(X_{j+\delta}^{\dagger}c_{j+\delta,\sigma}^{\dagger}c_{j\sigma}X_{j}+ {\rm H.c.})+\omega_{0}\sum_{j}
a_{j}^{\dagger}a_{j}\nonumber\\
&& -g^{2}\omega_{0}\sum_{j}n_{j}+(U-2g^2\omega_{0})\sum_{j}n_{j\uparrow}n_{j\downarrow} ,
\label{ai2}
\ee
where $X_{j}=e^{g(a_{j}-a_{j}^{\dagger})}$ and $n_j = n_{j \uparrow} + n_{j \downarrow}$.
Next, we express  as follows our LF transformed Hamiltonian in terms of the composite
fermionic operator
$d_{j\sigma}^{\dagger} \equiv c_{j\sigma}^{\dagger}X_{j}^{\dagger}$:
\be
H^{LF}_{hh}= -t\sum_{j\delta\sigma}\left(d_{j+\delta,\sigma}^{\dagger}d_{j\sigma}+
{\rm H.c.}\right)+\omega_{0}\sum_{j}
a_{j}^{\dagger}a_{j}\nonumber\\
+(U-2g^2\omega_{0})\sum_{j}
n_{j\uparrow}^{d}n_{j\downarrow}^{d}
-g^{2}\omega_{0}\sum_{j}\left(n_{j\uparrow}^{d}+n_{j\downarrow}^{d}\right) ,
\label{ai3}
\ee
where
 $ n_{j\sigma}^{d}= d_{j\sigma}^{\dagger}d_{j\sigma} $.
The last term is a constant polaronic energy and can be dropped. So Eq. (\ref{ai3}) 
essentially represents
the Hubbard Model for composite fermions
 with Hubbard interaction $U_{eff}=(U-2g^2\omega_0)$. The renormalization of Hubbard $U$
 by e-ph coupling has been recently observed in layered dichalcogenide $1$T-TaS$_2$\cite{insulating_1T_TaS}. In the limit of large $U_{eff}/t$,
using standard treatment involving a canonical (Hubbard to $t-J$) transformation,
we get the following effective Hamiltonian for the small parameter $t/U_{eff}$\cite{eder,troyer,bala}:
\be
H_{t-J}&=&P_s\left[ -t\sum_{j\sigma\delta}\left(d_{j+\delta,\sigma}^{\dagger}d_{j\sigma}+{\rm H.c.}\right)
+\omega_{0}\sum_{j}a_{j}^{\dagger}a_{j}\right. \nonumber\\
&+& \left. J\sum_{j\delta}\left(\vec{S}_{j}\cdot\vec{S}_{j+\delta}-{n_{j}^{d}n_{j+\delta}^{d}\over{4}}\right) \right] P_s
\label{ai5}
\ee
where $n_j^{d} =  n_{j\uparrow}^{d}+n_{j\downarrow}^{d}$,
$J={4t^2\over{U_{eff}}}$, $\vec{S}_j$ is the spin operator for a spin $1/2$
fermion at site $j$, and $P_{s}$ is the single-occupancy-subspace projection operator.
This is the $t-J$
Hamiltonian for the composite
fermionic operators $d_{j\sigma}$.

In terms of original operators $c_{j\sigma}$, the effective $t-J$ Hamiltonian in Eq. (\ref{ai5}) 
can be re-written as
\be
H_{t-J}=H_{0}+H_{1} ,
\label{ai7.0}
\ee
where
\be
H_{0}&=& -te^{-g^2}\sum_{j\sigma}P_s \left(c_{j+\delta,\sigma}^{\dagger}c_{j\sigma}+ {\rm H.c.} \right)P_s
+\omega_{0}\sum_{j}a_{j}^{\dagger}a_{j}\nonumber\\
&&+J\sum_{j}P_s\left(\vec{S}_{j}\cdot\vec{S}_{j+\delta}-{n_{j}n_{j+1}\over{4}}\right)P_s
\label{ai7.1}
\ee
and
\be
\!\!\!\!\!H_{1}&=& -te^{-g^2}\!\sum_{j\sigma}P_s \!\left[c_{j+\delta,\sigma}^{\dagger}c_{j\sigma}(Y_{+}^{j\dagger}
Y_{-}^{j}-1)+ {\rm H.c.}\right]\!P_s .
\label{ai7.2}
\ee

Here we have rewritten the above Hamiltonian to separate into (i) the electronic part $H_0$ which is nothing but
an effective $t-J$ model with reduced hopping ($t e^{-g^2}$); and (ii) the remaining perturbative part $H_1$ which corresponds
to the composite fermion terms containing the e-ph interaction with $Y^{j}_{\pm} \equiv e^{\pm g(a_{j+\delta}-a_{j})}$.

After carrying out perturbation theory to second-order (as outlined in Ref. \onlinecite{sryspbl1,sryspbl2}),
with $t/(g\omega_0 )$ as the small parameter \cite{pankaj_ys}),
we get the following effective Hamiltonian:
\be
H_{hh}^{eff}\cong -t_1h_{t_1} 
+J h_S -Vh_{nn} -t_2 h_{\s\s}-t_2 h_{\s \sb} \nonumber\\
\label{eff_ham}
\ee

{ where 
\be
h_{t_1} = \sd_{j\delta\sigma}P_s \left(c_{j+\delta,\sigma}^{\dagger}c_{j\sigma}+ {\rm H.c.}\right)P_s ,
\label{ht1}
\ee
\be
h_S = \sd_{j\delta}P_s \left
(\vec{S}_{j} \cdot \vec{S}_{j+\delta}-\frac{1}{4}n_{j}n_{j+\delta}\right)P_s ,
\label{hS}
\ee
\be
h_{nn} = \sd_{j\delta\s}(1- \!\!n_{j+\delta\sb})(1- \!n_{j\sb})(n_{j\s}- n_{j+\delta\s})^2 ,
\label{hnn}
\ee
\be
h_{\s \s} &&= \sd_{j \delta \delta^{'}\s}(1-n_{j+\delta^{'},\sb})(1-n_{j\sb})(1-n_{j+\delta,\sb}) \nonumber \\
&& ~~~~~~ \times \left[c_{j+\delta,\s}^\d(1-2n_{j\s})
c_{j+\delta^{'},\s}+ {\rm H.c.} \right] ,
\label{hss}
\ee
\be
h_{\s \sb} 
&& =\sd_{j\delta\delta^{'}\s}(1-n_{j+\delta,\sb})(1-n_{j+\delta^{'}\s}) \nonumber \\
&&~~~~~~\times \left[c_{j\s}^{\d}c_{j+\delta,\s}c_{j+\delta^{'},\sb}^\d c_{j\sb} + {\rm H.c.} \right] ,
\label{hss-}
\ee
The various coefficients are defined in terms of the system electron-phonon coupling $g$,
the Hubbard interaction $U$, the hopping amplitude $t$, and the phonon frequency $\omega_0$ as follows:  
$V\simeq t^{2}/{2g^2\omega_0}$, $J \equiv {4t^2\over{U-2g^2\omega_{0}}}$, $t_1 \equiv t e^{-g^2}$,
and $t_2\simeq t^2 e^{-g^2}/{g^2\omega_0}$. 
Hereafter $t=1$ is taken as unit of energy.

\begin{figure}[htp]
\centering
\includegraphics[width=.95\linewidth]{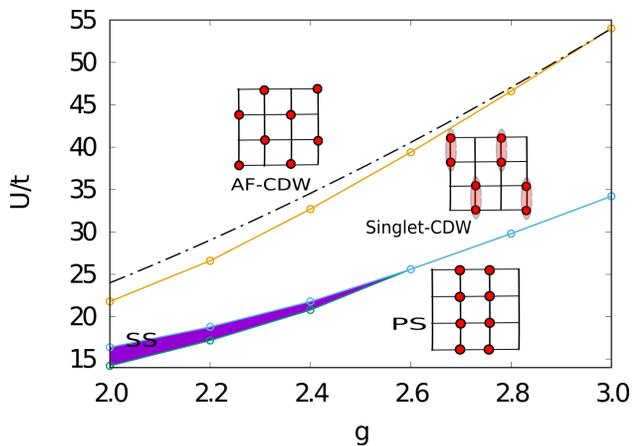}
\caption{
Different phases in $g-U/t$ plane. Phase separation (PS) i.e. antiferromagnetic clustering of electrons at smaller $U$ is broken to form insulating CDW made of NN singlets (two electron in shaded ellipse) i.e. singlet-CDW as we increase $U$. The breaking of singlet pairs happen with further increase in $U$ to give AF-CDW as shown. At smaller $g$, a metallic phase sandwiched between PS and singlet-CDW is stabilized in narrow range of parameters. The dashed black line is an etimate of the boundary between singlet-CDW and 
AF-CDW when hopping is ignored (see main text).}
\label{fig_phase_diag}
\end{figure}

Here we have the nearest neighbor parameter $\delta, \delta^{'}=(\hat{x},\hat{y})$ for each site $j$ to cover the 2D square lattice for the terms $h_{t_1},h_S, h_{nn}$ in effective Hamiltonian in Eq. \ref{eff_ham}. Here just to mention that the phonon averaging
(upto 2nd order perturbation) introduces a dominant NN repulsion term $h_{nn}$ which involves the electrons hopping to nearest neighbor sites and coming back. So this process prefers NN sites to be empty to avoid double occupancy and is basically a repulsion. Also this perturbation process includes longer range three site hopping process with further reduced amplitude $t_2$. For each site $j$ the next to NN (NNN) hopping terms $h_{\s \s}$ and $h_{\s \sb}$ have the sums over the parameters $(\delta,\delta^{'})=(\pm\hat{x},\pm\hat{y})$ to avoid double counting as described and shown in Fig.\ref{fig_model}. 

To study the ground state properties of this model on quasi-1D systems, we simulate the model on $4-$leg ladder system with periodic boundary condition in $y-$direction 
i.e., $4-$leg tube system by using DMRG method\cite{dmrg_white}. For different values of model parameters ($g, U/t$), 
we simulate the system with different tube length and calculate the ground state energies and order parameters. These physical 
quantities are then used to 
extrapolate the results to thermodynamic limit by finite size scaling. We keep upto $8000$ states of the density matrix to get the 
ground state energies with error $<10^{-8}$.

\section{Phase diagram}
Here we describe the different phases in $g-U/t$ plane at quarter-filling (one electron per two sites and 
equal number of up and down electrons) obtained by extensive DMRG simulation 
with $\omega_0/t=1$. As mentioned above, we have two dominant interaction terms in Hamiltonian: the effective Heisenberg interaction $J/t\equiv {4\over{U-2g^2}}$ which for a fixed $g$ decreases with the increase in $U$ and responsible for antiferromagnetic (AF) clustering of electrons i.e., phase separation (PS) and singlet pair formation; and an effective NN repulsion $V/t\simeq 1/({2g^2})$ which depends on $g$ only and tries to break Phase separated cluster of spins and singlet pairs. As shown in Fig.\ref{fig_phase_diag}, for large $g$ and smaller $U$, the system is phase separated i.e. clustering of spins happens due to large $J$. With increasing $U$, the cluster of spins breaks into singlet pairs (shaded pair of electrons) which at the 
commensurate quarter-filling are arranged in a insulating CDW state. We call it as singlet-CDW having structure factor peak at $S(\pi/2,\pi)$ or $S(\pi,\pi/2)$ depending on the orientation of the singlets. Further increase in $U$ decreases $J$ and the singlet pairs are broken to give another CDW phase as shown. The electrons are arranged in this phase to be AF order for smaller $g$ to gain some kinetic energy due to NNN hopping terms. 
The transition between phases is found to be abrupt which seems to be reasonable due to the negligible contribution of effective hopping terms for larger $g$. 

The situations is different for the case of smaller $g$ and smaller $U$ where NN hopping term $t_1=e^{-g^2}$ can be effective. Along with the CDW phases at larger $g$, we find a narrow range of metallic phase sandwiched between PS and singlet-CDW phase as shown as shaded area in Fig\ref{fig_phase_diag}. This phase is characterized by the vanishing charge gap, but with finite spin gap (singlet to triplet excitation). 
With increasing $g$, metallic phase is narrowed down and vanished at larger $g$.     

\begin{widetext}

\begin{figure}[htp]
\centering
\includegraphics[width=.99\textwidth]{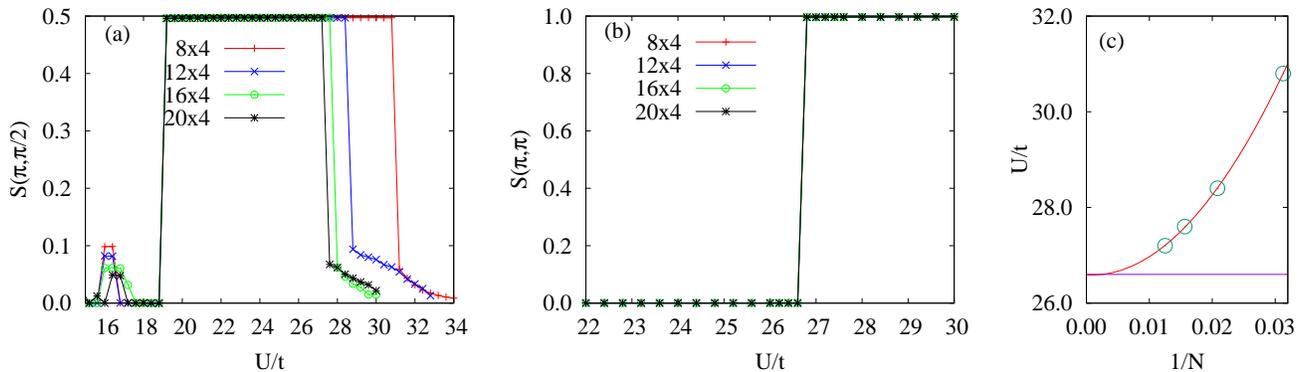}
\caption{The results are for $g=2.2$: (a) The order parameter $S(\pi,\pi /2)$ i.e. the structure factor for singlet-CDW phase when the edge potential is arranged as (${\delta}V, {\delta}V, -{\delta}V, -{\delta}V$) and (${-\delta}V, {-\delta}V, {\delta}V, {\delta}V$) on two edges of the tube system to pick up the phase. (b) The order parameter $S(\pi/2,\pi /2)$ i.e. the structure factor for AF-CDW when a edge potential is arranged as (${-\delta}V, {\delta}V, -{\delta}V, {\delta}V$) and  (${\delta}V, {-\delta}V, {\delta}V, -{\delta}V$) on two edges of the tube system to pick up the phase. (c) The finite size extrapolation to determine the phase boundary between singlet-CDW and AF-CDW.}
\label{fig_g2.2_struc_fac}
\end{figure}

\end{widetext}

\begin{figure}[htp]
\centering
\includegraphics[width=.98\linewidth]{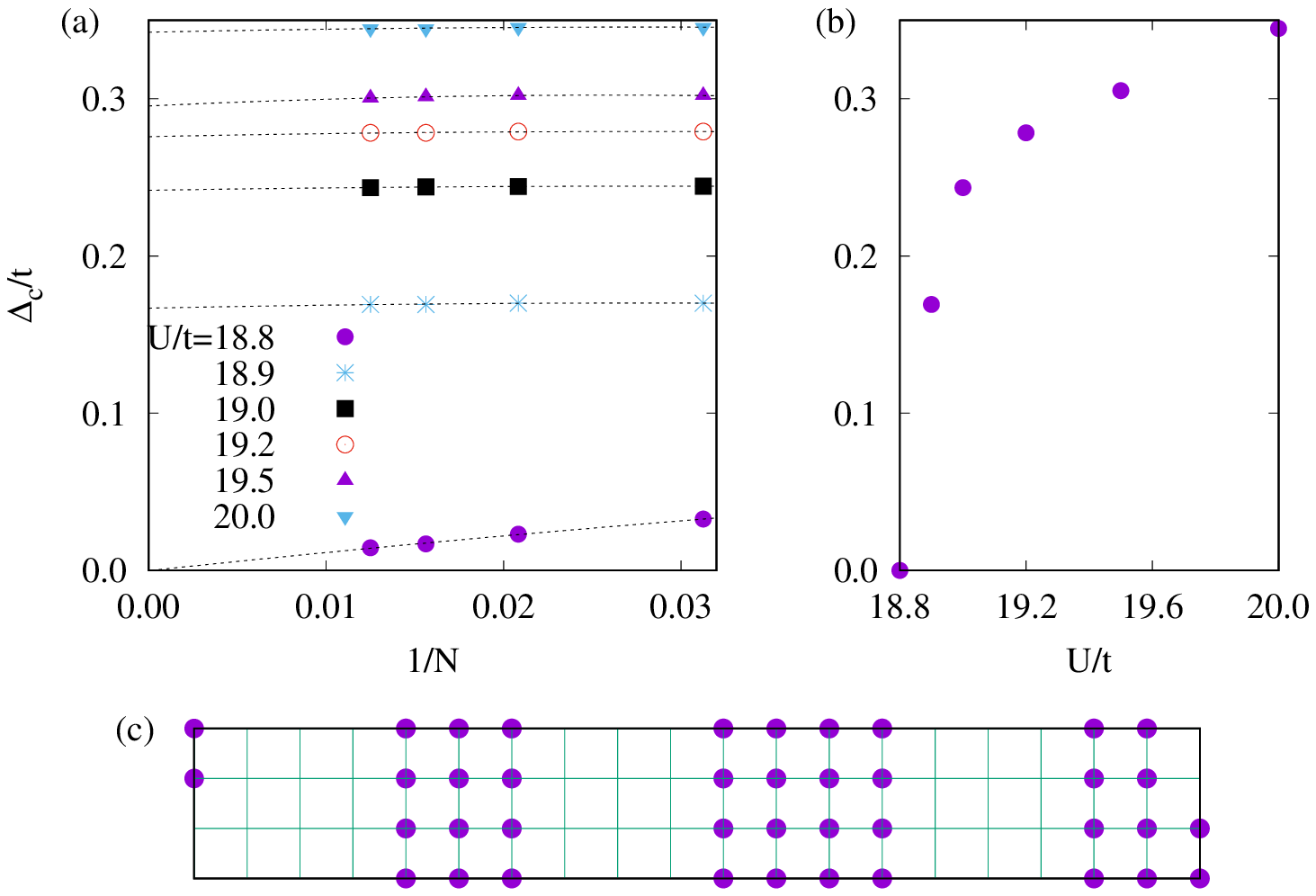}
\caption{ For $g=2.2$, (a) Typical extrapolation of charge gap ($\Delta_c$) approached from singlet-CDW phase to metallic phase. (b) $\Delta_c$ as a function of $U/t$ (c) Typical local density profile in phase separation.}
\label{fig_charge_gap_extr_1100}
\end{figure}

\section{Determination of Phase boundaries}
Here we discuss how we determine the phase boundaries in the phase diagram shown in Fig.\ref{fig_phase_diag}. The different insulating phases
are characterized by the structure factor peak which is the Fourier transform of density-density correlations and is defined as:
\be 
S({k_x},{k_y})=\frac{1}{N}\sd_{l_1,l_2}W(l_1,l_2)e^{-i(k_xl_1+k_yl_2)}
\ee  
where $W(l_1,l_2)=\langle n_{i,j}n_{i+l_1,j+l_2}\rangle$ is density-density correlation. For example, the singlet-CDW and AF-CDW phase can be captured by the structure factors $S(\pi,\pi /2)$ or $S(\pi/2,\pi)$ depending on the 
orientation of the NN singlets and $S(\pi/2,\pi /2)$ respectively.
The metallic phase is detected by vanishing charge gap defined as $\Delta_c=(E(N+2,0)+E(N-2,0)-2E(N,0))/2$ with $E(N,S_z^T)$ being the energy for $N$ number of electrons (equal up and down electrons) and total $z$ component of spins $S_z^T=0$. Also we confirm that metallic phase has non-zero spin gap $\Delta_s$ defined as $\Delta_s=E(N,1)-E(N,0)$. The phase separation has been captured by looking at the real space density profile obtained by DMRG simulations. 

We simulate 4-leg tube systems of different sizes i.e.,  $8$x$4$, $12$x$4$, $16$x$4$, $20$x$4$ systems with periodic boundary condition in y-direction (i.e., 4-leg tube) and open boundary in tube direction. We study the quarter-filled system i.e., number of particle is $N/2$ (equal up and down electrons) where $N$ is the tota number of sites. The physical quantities calculated for different system sizes enable us to extrapolate the results to thermodynamic limit by finite size scaling analysis as described below.

%

\subsection{Phase boundaries at smaller $g$}
In our simulation, we pick up the different CDW phases mentioned in the phase diagram by adding onsite potential at two edges of the tube. This is used to reduce the edge effect for these insulating phases and does not affect the results in thermodynamic limit. 

\begin{figure}[htp]
\centering
\includegraphics[width=.9\linewidth]{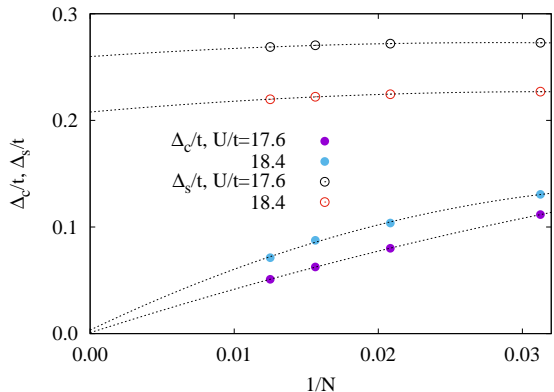}
\caption{ For $g=2.2$, typical extrapolation of charge gap ($\Delta_c$) and spin gap ($\Delta_s$) in the metallic phase.}
\label{fig_metallic_extrapolate}
\end{figure}

So, we use edge potential as (${\delta}V, {\delta}V, -{\delta}V, -{\delta}V$) on one edge and ($-{\delta}V, -{\delta}V, {\delta}V, {\delta}V$) on the other to pick up the singlet-CDW phase. Then we calculate the order parameter $S(\pi/2,\pi)$ for this phase in the intermediate region of the phase diagram. Note that the order parameter persists for arbitrary long system length and corresponds to long range singlet-CDW state. The results are shown for $g=2.2$ in Fig.\ref{fig_g2.2_struc_fac}(a). This shows finite size effect on singlet-CDW and AF-CDW phase boundary, but almost no effect on the boundary with PS phase. We also simulate the same systems after putting the edge potential as (${\delta}V, -{\delta}V, {\delta}V, -{\delta}V$) on one edge and ($-{\delta}V, {\delta}V, -{\delta}V, {\delta}V$) on the other to settle the AF-CDW boundary from above. The order parameter in this case is $S(\pi/2,\pi/2)$ and is shown in Fig.\ref{fig_g2.2_struc_fac}(b) to have no finite size effect. We then extrapolate the two results which are shown to coincide to same $U$ as shown in Fig.\ref{fig_g2.2_struc_fac}(c)--suggesting a abrupt transition between singlet-CDW and AF-CDW. This transition can also be captured analytically
ignoring the hopping contributions. The effective model contains $J$ and $V$ terms. So in singlet-CDW phase, the energy of a singlet corresponds to $-3J/4+(2V-J/4)$ which becomes zero at the singlet-CDW to AF-CDW transition points. After writing these in terms of $U$ and $g$, this gives the transition at $U/t\sim 6g^2$ which estimates 
the singlet-CDW to AF-CDW transition better at larger $g$ (see Fig.\ref{fig_phase_diag}). Also, we want to point out that we calculate the expectation value of the singlet operator 
$(S_i^{+}S_j^{-}+S_i^{-}S_j^{+})$ for each bond connecting two NN sites $i,j$ in singlet-CDW phase. This gives the value $-1$ (as it should be for a NN singlet pair between NN sites $i,j$) for each NN bond forming a NN singlet and $0$ otherwise.      

For smaller $g$, the NN hopping can be effective and we find that the breaking of cluster of spins at smaller $U$ goes through a metallic phase before forming singlet-CDW phase. 
To find the metallic phase boundary with singlet-CDW, we calculate charge gap $\Delta_c$
in the singlet-CDW phase with the edge potential as mentioned above. As we decrease $U$, we see the extrapolated $\Delta_c$ goes to zero as shown in Fig.\ref{fig_charge_gap_extr_1100}(a) and (b). The metallic boundary with phase separation has been captured by investigating the local electron density profile obtained by the DMRG simulation. The typical density profile in PS is shown in  Fig.\ref{fig_charge_gap_extr_1100}(c) where the filled circle size represents the total electron density. We also confirm the vanishing charge gap, but non-zero spin gap inside the metallic phase as shown in Fig.\ref{fig_metallic_extrapolate}. So this metallic phase is characterized as singlet superconducting (SS) phase as shown in Fig.\ref{fig_phase_diag}.

\begin{widetext}

\begin{figure}[htp]
\centering
\includegraphics[width=.99\textwidth]{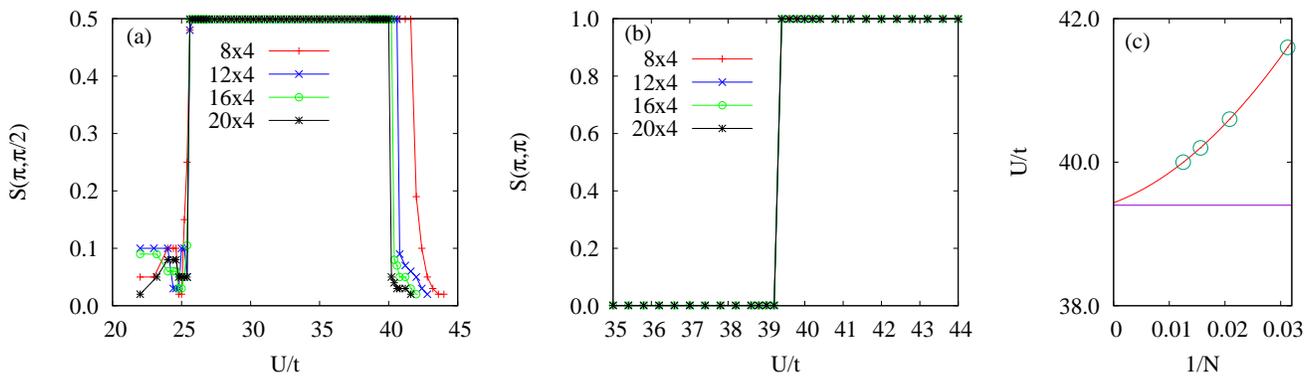}
\caption{Similar calculations for $g=2.6$ as shown in Fig.\ref{fig_g2.2_struc_fac}}
\label{fig_g2.6_struc_fac}
\end{figure}

\end{widetext}

\subsection{Phase boundaries at larger $g$}
Again we present the similar calculations for larger $g=2.6$ to detect the phase boundaries. The results are shown in Fig.\ref{fig_g2.6_struc_fac}. We see similar behavior at larger $U$ where the boundary approached from both CDW phases seems to be coinciding as shown in Fig.\ref{fig_g2.6_struc_fac}(c). In contrast to smaller $g$, we have not detected any metallic phase for $g=2.6$ and larger. This seems to be reasonable because for larger $g$, the hopping terms become less effective.
In Fig.\ref{fig_g2.6_charge_gap}, the typical extrapolation of charge gap ($\Delta_c$) around the singlet-CDW and PS boundary stays always finite. Although, a tendency to PS
for smaller system gives slightly negative $\Delta_c/t$, the 'normal' insulating state 
is restored for larger system sizes.

\begin{figure}[htp]
\centering
\includegraphics[width=.9\linewidth]{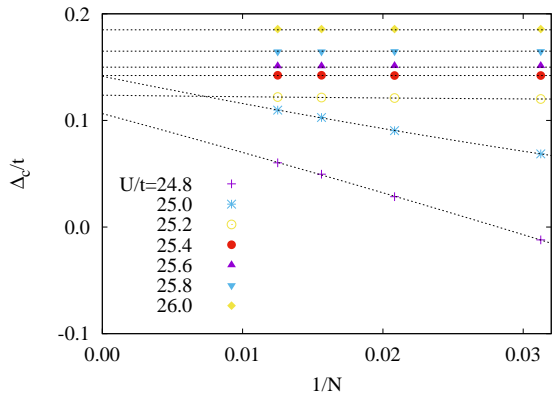}
\caption{ For $g=2.6$, typical extrapolation of charge gap ($\Delta_c$) around 
the singlet-CDW and PS bounday}
\label{fig_g2.6_charge_gap}
\end{figure}

\section{Conclusions}
In both strong electron-electron and electron-phonon coupling regime, we 
derive an effective electronic Hamiltonian for two dimensional Hubbard-Holstein model
by averaging out phonon degrees of freedom within second order perturbation theory. 
Using density matrix renormalization group method, we 
simulate the effective electronic Hamiltonian on $4-$leg tube systems to identify the different phases of the model
in $g-U/t$ parameter space. The phase boundaries are captured by structure factor peak, charge gap and real space density profile obtained from DMRG simulations which are extrapolated
to the thermodynamic limit.
We show that for larger $g$, the system goes through the phase separation, 
singlet-CDW and AF-CDW phases respectively as 
we increase $U/t$. The phase transitions seems to be abrupt as the effective hopping is negligible 
for larger $g$. For smaller $g$, we also get the similar CDW phases (AF-CDW and singlet-CDW) 
at larger $U/t$. But for smaller $U/t$, the hopping of electrons can be effective which 
gives rise to a metallic phase sandwiched between singlet-CDW and phase separation.
This phase is characterized by vanishing charge gap, but non-zero spin gap--suggesting a 
singlet superconducting phase. The phase diagram mostly contains the insulating phases.
In particular, the exotic singlet-CDW phase might be relevant to CDW phases arising from the interplay of electron-electron and electron-phonon coupling and observed in layered dichalcogenide $1$T-TaS$_2$\cite{insulating_1T_TaS}.    

\section{Acknowledgments}
S.R would like to thank Sudhakar Yarlagadda, Peter 
B. Littlewood and Herbert Fertig for stimulating discussions. 
We would like to thank U. Nitzsche for technical assistance. This work was supported by the NSF through
Grant Nos. DMR-1506263 and DMR-1506460 and by SFB 1143 of the Deutsche
Forschungsgemeinschaft. Computations were
carried out on the ITF/IFW Dreden, Germany.

\end{document}